\documentclass[12pt, onecolumn]{IEEEtran}
\usepackage{amsmath,epsfig}
\usepackage{amssymb}
\usepackage{amsfonts}

\usepackage{array,dcolumn}
\usepackage[ruled,vlined,linesnumbered]{algorithm2e}
\usepackage{psfrag}
\usepackage{graphicx}
\usepackage{amsmath}
\usepackage{amsfonts}
\usepackage{amssymb}
\usepackage{accents}
\usepackage{booktabs}
\usepackage{cite}
\usepackage{subfigure}
\usepackage[all]{xy}
\usepackage{dsfont}
\usepackage{url}
\usepackage{xcolor}
\usepackage[nolist]{acronym}

\begin{document}

\begin{acronym}
\acro{AWGN}{additive white Gaussian noise} \acro{EXIT}{extrinsic
information transfer}
\acro{RA}{random access}
\acro{CCw/oFB}{collision channel without
feedback}\acro{SIC}{successive interference
cancellation}\acro{SN}{slot node}\acro{UN}{user node}
\acro{CSA}{coded
slotted aloha}
\acro{MAC}{medium access control}\acro{MAP}{maximum
a-posteriori}\acro{CRA}{coded random access}\acro{SPC}{single
parity-check}\acro{MDS}{maximum distance
separable}\acro{RS}{Reed-Solomon}\acro{SNR}{signal-to-noise
ratio}\acro{D-GLDPC}{doubly-generalized low-density
parity-check}
\acro{DE}{differential evolution}
\acro{PLR}{packet loss rate}
\acro{SA}{slotted ALOHA}
\acrodef{IRSA}{irregular repetition slotted ALOHA}
\acrodef{LDGM}{low-density generator-matrix}
\acrodef{LT}{Luby-transform}
\acrodef{LDPC}{low-density parity-check}
\acrodef{BEC}{binary erasure channel}
\acrodef{p.m.f.}{probability mass function}
\end{acronym}

\title{Coded Random Access: Applying Codes on Graphs to Design Random Access Protocols}
\author{\IEEEauthorblockN{Enrico Paolini}
\IEEEauthorblockA{\\ Department of Electrical, Electronic, and Information Engineering ``G. Marconi'',\\ University of Bologna, Italy\\
e.paolini@unibo.it
}\\[2mm]
\and
\IEEEauthorblockN{\v{C}edomir Stefanovi\'{c}}
\IEEEauthorblockA{\\Department of Electronic Systems, Aalborg University, Denmark\\ Email: cs@es.aau.dk }\\[2mm]
\and
\IEEEauthorblockN{Gianluigi Liva}
\IEEEauthorblockA{\\Institute of Communications and Navigation, DLR, Germany\\
Email: gianluigi.liva@dlr.de}\\[2mm]
\IEEEauthorblockN{Petar Popovski}
\IEEEauthorblockA{\\Department of Electronic Systems, Aalborg University, Denmark \\
Email: petarp@es.aau.dk}
\thanks{The work of E. Paolini was supported by the EU's Seventh Framework Programme (FP7/2007-2013) under grant agreement n. 288502. The work of \v C. Stefanovi\' c was supported by Danish Council for Independent Research, grant no.  DFF-4005-00281. The work of P. Popovski was partially supported by the Danish Council for Independent Research, Sapere Aude Grant No. 11-105159.}
}

\maketitle

\begin{abstract}
The rise of machine-to-machine communications has rekindled the interest in random access protocols as a support for a massive number of uncoordinatedly transmitting devices.
The legacy ALOHA approach is developed under a collision model, where slots containing collided packets are considered as waste. However, if the common receiver (e.g., base station) is capable to store the collision slots and use them in a transmission recovery process based on successive interference cancellation, the design space for access protocols is radically expanded. We present the paradigm of \emph{coded random access}, in which the structure of the access protocol can be mapped to a structure of an erasure-correcting code defined on graph. This opens the possibility to use coding theory and tools for designing efficient random access protocols, offering markedly better performance than ALOHA. Several instances of coded random access protocols are described, as well as a case study on how to upgrade a legacy ALOHA system using the ideas of coded random access.
\end{abstract}

\section{Introduction}
\label{sec:Intro}

We start with a deceptively simple question: When and why should we use random access? A concise answer would be: Whenever there is an uncertainty about the set of users that aim to transmit at a given instant.
A canonical scenario falling in the above description is the one in which a set of uncoordinated devices aims to transmit over the shared wireless medium to the same receiver at approximately the same time, and the random access mechanisms are needed to break this ``symmetry'' and enable successful access.
As such, random access is an essential component of any distributed wireless communication system, typically used for initial link establishment or distributed spectrum sharing among interfering networks, such as two collocated WiFi hotspots.
Presently, we are witnessing a revival of research interest in random access mechanisms, driven by the increasing presence of \emph{machine-to-machine (M2M)} communications in cellular and satellite networks.
Efficient random access is instrumental in M2M scenarios, due to the fact that there is a massive and uncoordinated set of transmitting devices.

ALOHA \cite{R1975} is a rather generic form of random access, typically operating under the assumption that collided packets are irrecoverably lost.
Standard variants of the ALOHA protocol aim to maximize the number of collision-free transmissions within a given time interval, i.e., to maximize the expected throughput.
In slotted ALOHA (SA) \cite{R1975}, link time is divided into equal-duration slots, and the devices are slot-synchronized, contending for access on a slot basis with a predefined slot-access probability.
A related solution is framed slotted ALOHA (FSA) \cite{OIN1977}, where slots are organized into frames, and the users transmit in a single, randomly chosen slot of the frame.
In both variants, only the slots containing a single transmission (singleton slots) are useful and the corresponding transmission is successfully received, while the slots containing no transmission (idle slots) or multiple user transmissions (collision slots) are wasted.
The throughput $T$, defined as the probability of successfully receiving a user transmission per slot, is equal to the probability that a singleton slot occurs.
The maximal asymptotic throughput in both variants is a rather low $T_{\max} = 1/e \approx 0.37$.

Recently there has been a conceptual shift in the theory and practice of slotted ALOHA protocol family, based on the use of successive interference cancellation (SIC) that enables ``unlocking'' of the collisions slots. Some of these advances apply SIC at slot level, in order to separate the collided signals and allow multiple packets being received within a single slot, c.f. \cite{ZZ2012},  which may be regarded as an instance of multi-user detection (MUD). These access protocols, applied also to combat the hidden terminal problem in carrier sensing multiple access (CSMA) systems \cite{TDN2011}, still rely on an instantaneous feedback from the receiver, notifying the transmitters about unresolved collisions and initiating retransmissions. Other recent advances consist of combining SA with physical layer network coding \cite{GGW2013}.

This work is dedicated to a conceptually different improvement, based on SIC across multiple slots \cite{CGH2007}. The essence of these modifications is rather simple: active devices transmit replicas of the same packet in multiple slots, while SIC is used on the receiving side to remove replicas of already recovered transmissions from collision slots. Recovery and removal of replicas is performed in an iterative, i.e., successive manner, where new iterations are propelled by the transmissions recovered in the previous round, as illustrated in Fig.~\ref{fig:SIC}. The exploitation of the collision slots boosts the throughput -- in a basic scenario where active devices transmit two replicas in randomly selected slots of a frame \cite{CGH2007}, the asymptotic throughput increases to $T_{\max} \approx 0.55$.\footnote{One may argue that the comparison with standard FSA is unfair, as in FSA a user sends only one packet replica before receiving feedback on the contention outcome. However, it should be noted that in standard FSA a user may also transmit multiple replicas in order to get the data through, the difference is that the retransmission is initiated by the feedback.} The true potential of the SIC-enabled slotted ALOHA was revealed in \cite{L2011}, identifying analogies with modern channel coding based on sparse graphs and establishing the paradigm of \emph{coded random access}. The objective of this paper is to introduce these new developments, identify the ways in which they can be beneficial for M2M applications and highlight the important implementation issues. The outlined concepts are applicable in all systems which exploit slotted ALOHA, e.g., in random access channels of the cellular access and of the next generation interactive satellite services.

\section{Basics of Coded Random Access}
\label{sec:Basics}

\subsection{Access Scheme Description}\label{sec:access_scheme_description}

We start by considering \ac{CSA} in which the access is organized in \emph{contention periods}. Each contention period is a frame containing $M$ slots of equal duration, where $M$ is fixed. A set of $N$ users uses contention periods to communicate with a Base Station (BS), which acts as a common receiver. We are interested in the regime where the user population is large with respect to the size of the contention period $N\gg M$, but only a subset $N_{\mathsf a}$ of the users is active in a given contention period.
A simple model to create the uncertainty in the set of active users can be described as follows. At the beginning of a contention period each user independently generates a packet to be transmitted with \emph{activation probability} $p_{\mathsf a}$, where $p_{\mathsf a} \ll 1$. 
The number of active users in a contention period $N_{\mathsf a}$ is then a binomially distributed random variable, with mean value $\bar{N}_{\mathsf a}=p_{\mathsf a} N$.

\begin{figure}[h]
	\begin{center}
\includegraphics[width=0.35\columnwidth]{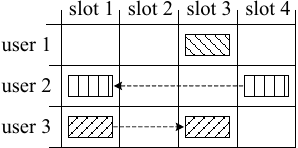}
	\end{center}
\caption{Successive interference cancellation in slotted ALOHA. Packet of user 2 is recovered in slot 4, enabling the recovery of packet of user 3 in slot 1, performed by subtracting the replica of user 2 packet in slot 1. In the same way, recovery of packet of user 3 enables the removal of its replica from slot 3, thus recovering packet of user 1. In this example, the use of SIC grants throughput of 0.75~packet/slot; without SIC, the throughput drops to 0.25~packet/slot.}
	\label{fig:SIC}
\end{figure}

The \ac{CSA} scheme works as follows.
Each active user generates $d$ packet replicas, where the repetition rate $d$ is drawn randomly according to a pre-determined probability distribution, which is the same for all users. The repetition rate is picked by an active user independently of all other active users and independently of all his previous choices.
The $d$ replicas are then transmitted by the user over $d$ slots picked uniformly at random among the $M$ slots of the contention period.
Following the example of Fig.~\ref{fig:SIC}, the users $2$ and $3$ picked a repetition rate $d=2$, while user $1$ did not replicate its packet, i.e., its repetition rate is $d=1$.
A packet in a singleton slot is decoded correctly. Each packet is assumed to contain pointers to describe the positions of the other replicas in the contention period sent by the same user\footnote{An efficient way to transport pointers is discussed in Section~\ref{sec:issues}.}. The packet is then re-encoded and re-modulated and the receiver removes its interference contribution from the $d-1$ slots containing the replicas.
The process proceeds iteratively, i.e., recovered replicas may lead to solving other collisions, as illustrated in Fig.~\ref{fig:SIC}.

The \emph{rate} of the the \ac{CSA} scheme is defined as
\begin{align}\label{eq:rate}
R=\frac{1}{\bar{d}}
\end{align}
where $\bar{d}$ is the average number of replicas sent per user. Obviously, a lower rate implies higher number of repetitions and the use of more energy per useful bit. The \emph{logical} load of the channel is defined as the expected number of active users per slot,
\begin{align}\label{eq:average_G}
G =\frac{\bar{N}_{\mathsf a}}{M}=p_{\mathsf a} \frac{N}{M} ,
\end{align}
i.e., the logical load corresponds to the expected number of new packets generated during the contention period.
The \emph{physical} load of the channel, i.e., the expected number of all transmitted replicas, is given by $G_{\mathsf{phy}}=G \cdot \bar{d}$.
In standard FSA there is only a single replica $d=1$, thus the logical and physical loads coincide.

\subsection{Bipartite Graph Representation and Asymptotic Analysis Over a Collision Channel}
\label{sec:bi}

\begin{figure}[h]
\begin{center}
\includegraphics[width=0.8\columnwidth,draft=false]{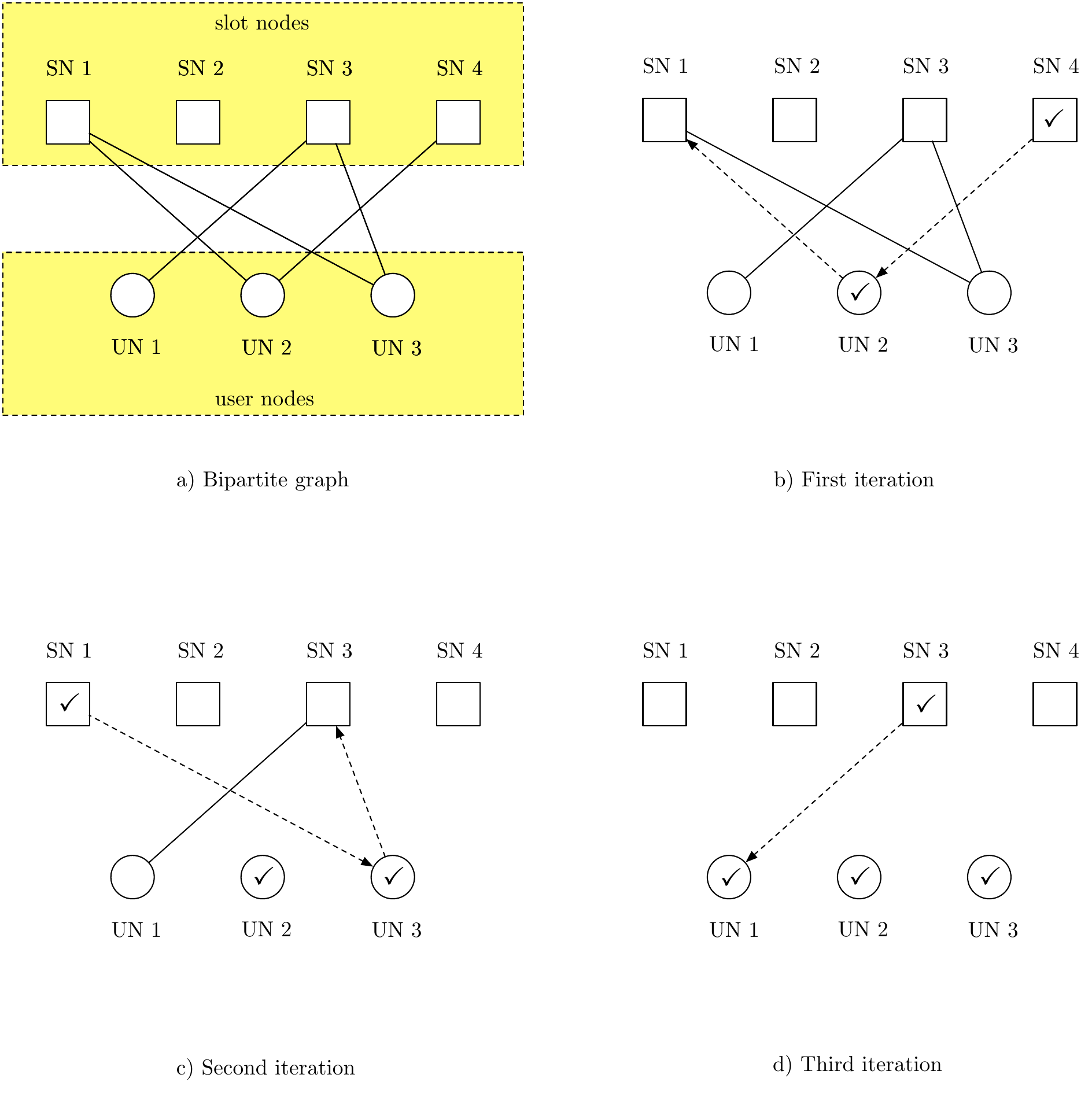}
\end{center}
\caption{Bipartite graph representation of the access scheme of Fig.~\ref{fig:SIC}.}\label{fig:CRA_Graph}
\end{figure}

Fig.~\ref{fig:CRA_Graph} shows the graph representation of the \ac{CSA} scheme for the example on Fig. \ref{fig:SIC}.
Specifically, it is represented by a bipartite graph, consisting of a set $N_{\mathsf a}$ \emph{user nodes} (one for each active user), a set of $M$ \emph{slot
nodes} (one for each slot), and a set of edges.
An edge connects the $i$th \ac{UN} the $j$th  \ac{SN} if and only if the user $i$ sends a packet in the $j$th slot.
The \emph{degree} $d$ of a given \ac{UN} is equal to the number of edges connected to it, each edge corresponding to one of the replicas sent by the user.
This graphical representation allows to establish a connection between the SIC procedure and iterative decoding of channel codes based on sparse graphs. This connection is here illustrated under the following assumptions:
\begin{enumerate}
\item For each slot, the receiver always discriminates between a ``silence'', singleton or a collision.
\item When a packet is received in a singleton slot, data are always correctly decoded.
\item Channel estimation and the interference cancellation are ideal.
\end{enumerate}
The first two assumptions are typical for \emph{collision channel} models.
The channel estimation has to be performed to enable SIC and the third assumption simplifies the analysis without substantially affecting the obtained results, as shown in \cite{CGH2007}.
Further, error events due to fading and thermal noise may affect the performance; in this regard, the reader may refer to \cite{CGH2007,L2011}.
We also outline the main difference to codes on graphs: the degree of a slot node cannot be controlled and it can even be equal to zero (idle slot). Clearly, if the BS could control the degree of each slot, we would not need random access at all, as a single user would be scheduled in each slot.

Under the above assumptions, the SIC procedure may be described as an instance of the iterative \emph{peeling decoder} for codes constructed on sparse graphs and transmitted over a \ac{BEC} \cite{LMS1998}.
The decoder consists of initializing the status of all \acp{UN} to ``unknown'' and of repeating the following procedure until the status of all \acp{UN} has been updated to ``known'', in which case decoding terminates successfully, or until at some iteration the status of no \ac{UN} is updated, when a failure is declared. The procedure is described as follows:
\begin{itemize}
\item For all \acp{SN}, if the \ac{SN} has degree $1$ then update to ``known'' the status of the unique \ac{UN} connected to it.
\item Remove all edges connected to the \ac{UN} and update the degrees of the \acp{SN} accordingly.
\end{itemize}
The way the SIC mimics the peeling decoder is illustrated in Fig. \ref{fig:CRA_Graph}b)-d).

The analogy between SIC for \ac{CSA} and iterative decoding of codes on sparse graphs allows to use techniques developed in the field of coding theory and apply them to random access. Accordingly, collisions are favored by \ac{CSA}, in a statistically controlled manner. For example, the theory of codes on graphs allows to properly design the probability distribution with which the users select their degrees to generate bipartite graphs on which SIC is successful with high probability. Judiciously designed probability distributions yield irregular graphs favoring the SIC procedure. Moreover, through the application of analytical tools from the theory of codes on graphs, such as density evolution or \ac{EXIT} charts, we can show the existence of a thresholding behavior of \ac{CSA} under SIC. This happens when both the frame size $M$ and the user population size $N$ tend to infinity, but the ratio $\frac{N}{M}$ remains constant. It turns out that there exists a threshold value
$G^*$, such that when the logical load is $G \leq G^*$, the SIC procedure almost certainly terminates successfully, i.e., each active user manages to send the packet to the BS within the contention period. Conversely, if $G > G^*$ then the opposite is true, i.e., there is a fraction of users' packets which will certainly not be decoded. It is possible to show that the threshold $G^*$ depends both on the selected user rates and on the probabilities with which these rates are selected. With a suitable selection of the repetition rates and their associated probability distribution, a threshold as large as $G^*=1$ packet/slot can be achieved. In other words, the throughput performance becomes equivalent to the perfectly scheduled access! The way the rate distribution is optimized follows the footsteps of the degree distribution optimization algorithms used in the design of \ac{LDPC} codes \cite{PLC2011b}.

As both the threshold $G^*$ and the rate $R$ are functions of the repetition rates distribution, one may look for the maximum achievable threshold $G^*$ for a given rate $R$. Note that when repetition coding is used, the rate is necessarily $0 < R \leq 1/2$, as there are at least two repetitions. Once $R$, as defined in \eqref{eq:rate}, is fixed, it can be shown that the  threshold ${G}^*$ of a \ac{CSA} scheme is upper bounded by the unique positive real solution of the equation
\begin{align}
\label{eq:CSA_cap_bound}
G=1-e^{-G/R} \, ,
\end{align}
as shown in \cite{PLC2011b}. If the user invests more power by increasing the number of repetitions, then $R$ decreases and the right-hand side of (\ref{eq:CSA_cap_bound}) increases, also implying that the upper bound increases.

\section{Variants of CSA}
\label{sec:Variants}

\subsection{High-Rate \ac{CSA} from Generic Component Codes}

The upper bound resulting from (\ref{eq:CSA_cap_bound}) is valid for every rate $R$ between $0$ and $1$.
In order to achieve rates $R > 1/2$, \cite{PLC2011b} introduces a generalization of the \ac{CSA} protocol that uses generic linear block codes instead of repetition codes. In this setting, a user that is active in a given contention period, splits his packet into $k$ \emph{segments} of the same length.
The $k$ segments are then encoded using a linear block code and $d$ segments are obtained as output. The linear block code is drawn randomly by the user from a set of component codes, according to pre-determined probability distribution. The information about the code used to encode the $k$ segments may be conveyed in a header appended to each segment. The component codes may have different lengths $d$, but they all have the same dimension $k$.
The rate of this generalized scheme is given by $R = k / \bar{d}$, where $\bar{d}$ is the expected length of the employed component code. This definition of the rate coincides with that given in \eqref{eq:rate} when repetition codes are used. With a judicious selection of $k$, of the lengths $d$ of the component codes and of their probability distribution, any rate $0 < R < 1$  can be obtained. Note that the choice $k=1$ reduces this generalized framework to the repetition-based case.

The $d$ encoded segments, equipped with appropriate pointers in their headers, are transmitted over $d$ slots picked uniformly at random within the contention period.
The contention period is now organized into $k M$ slots, each of the same time duration as that of a segment; the time duration of the contention period is thus the same as in the repetition-based case.
The bipartite graph representing the access scheme is now composed of $kM$ \acp{SN} and $N_{\mathsf a}$ \acp{UN}, where now each \ac{UN} corresponds to $k$ segments.
On the receiver side SIC is performed similarly to the repetition-based case, the only difference being the execution of some form of erasure decoding at the generalized \acp{UN} at each iteration.
In case simple codes are used, \ac{MAP} erasure decoding may performed.
Similar to the case with repetition, thresholding phenomenon is also observed for the high-rate \ac{CSA}.

\subsection{Spatially Coupled \ac{CSA}}

A variant of the \ac{CSA} scheme, is based on \emph{spatial coupling}, a technique widely used in the field of modern error correcting codes.
We present it in a simplified scenario in which all users exploit the same packet repetition rate $d$.

In the \emph{spatially coupled CSA}, a user becoming active at the beginning of a contention period with $M$ slots is allowed to transmit only one replica in that period, as opposed to the scheme described in Section~\ref{sec:access_scheme_description} in which all $d$ replicas are transmitted in that contention period.
Each of the other $d-1$ replicas is transmitted by the user in one of the subsequent $d-1$ periods.
Assuming the average number of active users per contention period is $\bar{N}_{\mathsf a}=p_{\mathsf a} N$, on average there are $p_a N$ packet replicas in the first contention period (one per active user), $2 p_{\mathsf a} N$ packet replicas in the second contention period (one per user becoming active at the beginning of the first period and one per user becoming active at the beginning of the second period), etc. up to the $d$-th contention period in which we expect $d p_{\mathsf a} N$ packet replicas on average. The expected number of replicas in a contention period that comes after the $d-$th one ``stabilizes'' to $d p_{\mathsf a} N$.
Thus the expected physical load is $G_{\mathsf{phy},1}=G$ in the first contention period, see \eqref{eq:average_G}, then it is $G_{\mathsf{phy},2}=2G$ in the second contention period, etc., and stays $G_{\mathsf{phy},d}=d\,G$ from the $d-$th period and onwards.

As shown \cite{LPLC2012}, the probability of a collision in a slot that belongs to a given contention period increases with the physical load imposed on that period. Due to the lighter physical load, the first contention period contains a lower number of collisions. The packets received in singleton slots of the first contention period may be used to remove the contribution of interference of their replicas in all $d-1$ subsequent contention periods. Therefore, although a slightly higher number of collisions are expected in the second contention period, some of them are resolved by interference cancellation. The resolved collisions are exploited, together with the packets received in the singleton slots from the first and second periods, to resolve further collisions in the third period. This process, when iterated through the sequence of contention periods, determines a ``chain reaction'' which allows to resolve more collisions than those resolved by the scheme in Section~\ref{sec:access_scheme_description} for the same repetition rates and probability distribution. Moreover, a thresholding phenomenon is again observed.
Specifically, the iterative decoding threshold of the spatially coupled scheme reaches the theoretical, upper-bound threshold of the block scheme under optimal, MAP decoding on a priori known graph\footnote{We again stress the fact that in CSA the graph is not known a priori due to the randomness of the contention process.}!

\subsection{Frameless CSA}

Finally, we introduce \emph{frameless} ALOHA \cite{SPV2012}, a variant of the CSA scheme inspired by the rateless codes \cite{BLMR1998}.
Two essential differences to the previously described CSA protocols are:
\begin{itemize}
\item When the contention period starts, the active users decide whether or not to transmit on a slot basis, as the slots ``appear'' on the wireless medium.
\item The contention period duration is not a-priori determined, but it is adaptive and tuned to the evolution of the contention/packet-recovery process.
\end{itemize}
In general case, both the user access strategy (i.e., the choice of slot-access probabilities) and the contention termination criterion are subject to optimization.
In \cite{SP2013} a simple version of the scheme was investigated, where the access strategy is ``memoryless'' and the slot-access probabilities are uniform both over users and slots. The scheme uses a heuristic termination criterion: the receiver monitors both the instantaneous throughput and the fraction of resolved user packets and, when either of them surpasses a predefined threshold, the contention is terminated through a suitable feedback signal.
It was shown that, although asymptotically suboptimal, this approach grants throughputs that are the highest in the reported literature for low to moderate number of active users, i.e., when $N_{\mathsf a}$ in the range $50 - 1000$.

\begin{figure}[h]
	\begin{center}
\includegraphics[width=0.75\columnwidth]{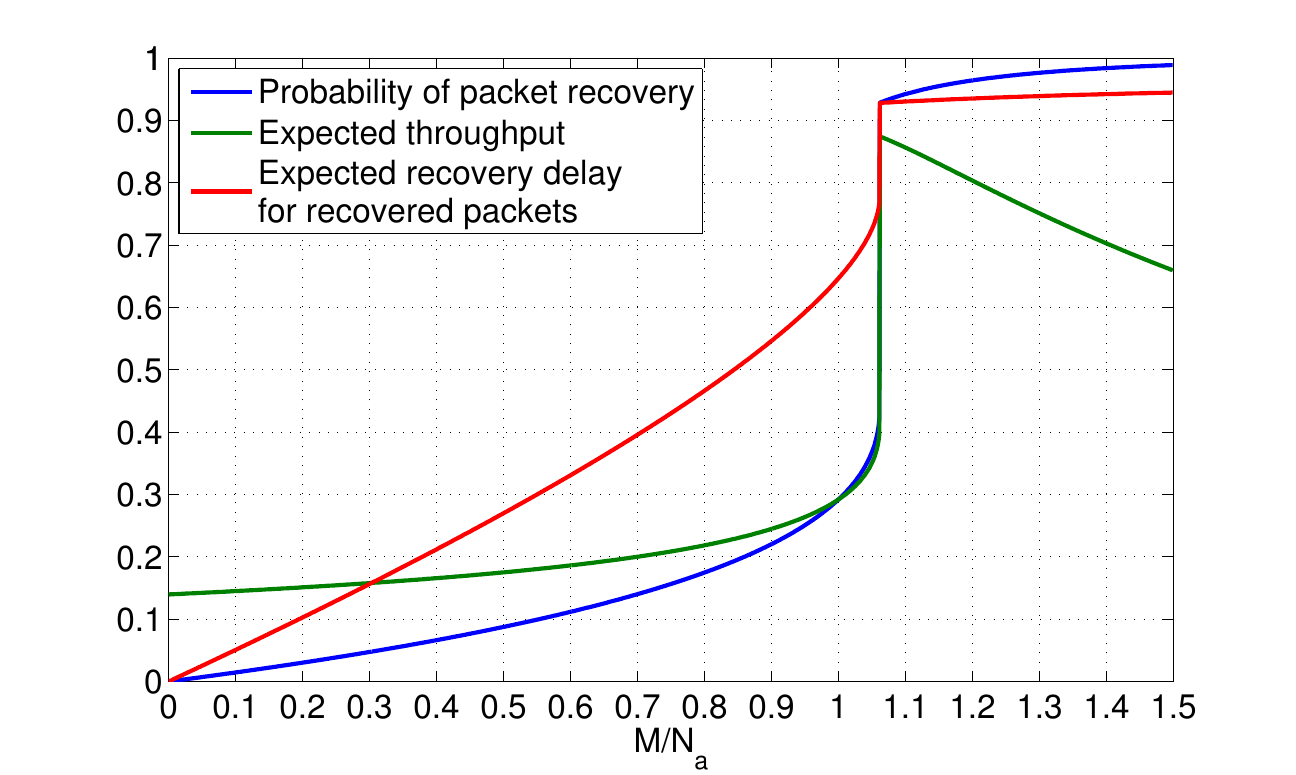}
	\end{center}
\caption{Asymptotic performance of frameless CSA.}
	\label{fig:frameless}
\end{figure}

Fig.~\ref{fig:frameless} illustrates the asymptotic performance of frameless ALOHA, showing the probability of packet recovery, expected throughput and expected recovery delay of recovered packets, as functions of the number of elapsed slots vs the number of active users $M / N_{\mathsf a}$.
The slot-access probability in the example is set to $3.1 / N_{\mathsf a}$, a value that maximises the expected throughput \cite{SPV2012}.
It is seen that the probability of packet recovery at first increases slowly and then rises steeply for some critical value of $M / N_{\mathsf a}$. The same behavior is also observed in iterative BP erasure-decoding of rateless codes.
The critical $M / N_{\mathsf a}$ actually defines the (expected) asymptotically optimal length of the contention period with respect to the throughput maximization, also observed in Fig.~\ref{fig:frameless}.
Finally, the expected recovery delay for recovered packets increases linearly until the critical $M / N_{\mathsf a}$. Although this behavior seems favorable, one should take into account that most of the packets are actually not recovered and thus do not contribute to the calculation of the delay.
After critical $M / N_{\mathsf a}$, most of the packets become recovered and the delay saturates.

The principle of adaptive termination favors the ``fortunate'' instances of packet-recovery process, ending the contention as soon as the terminating conditions are met \cite{SP2013}.
The adaptive termination also implies that the packet-recovery process can tune to the actual wireless link conditions and potential imperfect SIC instances, simply disregarding the affected slots and proceeding with the contention process.
In other words, frameless CSA is inherently adaptable to the scenarios when the assumptions outlined in Section~\ref{sec:bi} may not hold.
The main drawback is that the moment when the users receive feedback that terminates the contention is not known a-priori.
In scenarios where the uplink and downlink transmissions share the same spectrum, in frameless CSA the BS has to contend with the active users when transmitting the feedback, as analyzed in \cite{SP2013}.
We conclude by noting that similar arguments apply when comparing the advantages/drawbacks of the block and rateless coding frameworks.

\section{Practical Issues}
\label{sec:issues}

One of the underpinning assumptions of CSA is that each replica is equipped with pointers to the slots containing other replicas transmitted by the same user.
However, in practice, it is neither trivial to make the pointers nor the cost of sending many pointers is negligible.
A more elegant approach to address this issue is to embed in each replica a user-specific seed of a pseudorandom generator known both to the user and the BS.
Once a replica is resolved, the BS can use the knowledge of the generator and the obtained seed to determine all the slots containing the other replicas.

Another important practical issue is the estimation of the number of active users in a contention period $N_{\mathsf a}$, which is usually a priori not known and may vary over time, and which is required both in the framed and frameless variants of CSA in order to attain the optimal performance.
Specifically, in framed CSA the knowledge of $N_{\mathsf a}$ should be used to dynamically adapt the duration of the contention period size $M$, in order to guarantee a constant logical load and thus a constant throughput. In frameless CSA, both the optimal slot-access probabilities and the termination criterion depend on $N_{\mathsf a}$ \cite{SP2013}. An efficient estimation algorithm specifically tailored for frameless version of the scheme was proposed in \cite{STPP2013}.

\section{Case Study: Upgrading the Existing Slotted ALOHA Implementations}

Coded random access protocols can be very useful in the context of M2M communications, both in cellular and satellite access. 
Specifically, the access reservation procedure in all cellular standards, from GSM, over 3G, to LTE, is commonly based on the framed slotted ALOHA, providing acceptable performance for human-oriented traffic. However, the M2M traffic has fundamentally different requirements, primarily seen in the massive number of accessing terminals with short reporting deadlines, and the traditional ALOHA may create bottlenecks already in the access reservation.

We present a short study, describing how the contention phase of an existing cellular access reservation protocol can be upgraded to reap the advantages of coded random access while preserving the physical-layer behavior of the devices unchanged. The required modifications on the device side could be reduced to the implementation of the pseudorandom generators that will drive the selection of slots in which the access will be performed. This includes a downlink signaling between the BS and the devices, in order to tune the pseudo-random generators, timers, back-off exponents and other parameters of the actual FSA implementation, c.f. \cite{3GPPTS36.321}.
On the other hand, the BS stores the received uplink signals and uses SIC to process them, thereby absorbing the complexity of the upgrade, which is another highly desirable feature in practice.

\begin{figure}[h]
	\begin{center}
\includegraphics[width=0.95\columnwidth]{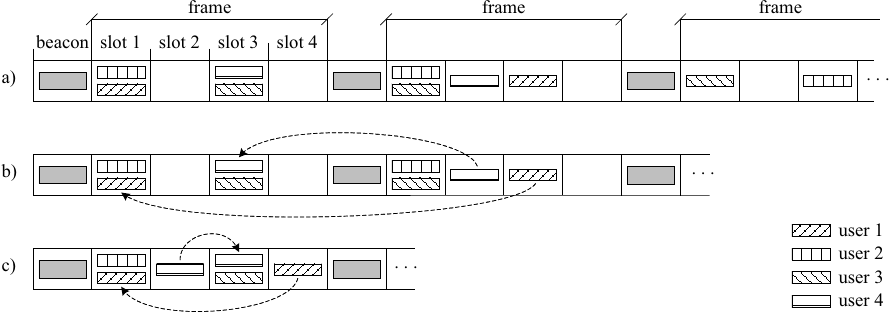}
	\end{center}
\caption{Example upgrade of framed slotted ALOHA.}
	\label{fig:SA_example}
\end{figure}

Fig.~\ref{fig:SA_example}a) presents an example of a generic framed slotted ALOHA. Active users transmit just once per frame and only the transmissions occurring in singleton slots are successfully received and the corresponding devices are notified via the next beacon. The unsuccessful ones continue transmitting in the subsequent frames, choosing the slots where the repeated transmission take place independently with respect to the choice made in the previous frames. In the example, packets of user 1 and user 4 get through in the second frame, and of user 2 and user 3 in the third frame.

In a simple upgrade, Fig.~\ref{fig:SA_example}b), the active users also transmit once per frame, as in typical FSA. Nevertheless, the slot choice is dictated using the CSA approach, modified such that there can be only a single transmission within the subset of slots that belong to a frame. This effectively translates to a constraint imposed on the possible edge configurations in the bipartite graph. The slot choice is made locally at each user using a predefined function derived through the CSA graph-based design, whose inputs are the user ID and the information received from the beacons sent by the BS. Once a transmission is recovered, the BS retrieves the corresponding user ID, which enables the backtrack and cancellation of the replicas from the previous frames and potential resolution of other transmissions. In the example from Fig.~\ref{fig:SA_example}b), the recovery of packet of user 1 and user 4 in the second frame allows to recover packets of user 2 and user 3 from the first frame; for the sake of simplicity, we assumed that choice of the slots is the same as in Fig.~\ref{fig:SA_example}a).

Finally, the full upgrade that matches the standard CSA is presented in Fig.~\ref{fig:SA_example}c). The users are allowed to repeat the same transmissions in multiple slots of the frame; the access strategies are again determined locally according to a predefined function, derived through the CSA approach and depending on the user ID and the information received from the BS. In this case, the BS removes the recovered packets both in ``forward'' and ``reverse'' directions.

We conclude by noting that the application of the concepts described above could be made both in protocols that contend with data and protocols based on access reservation.

\section{Conclusion}

The legacy slotted ALOHA, although essentially inefficient, underpins the majority of the existing wireless random access protocols.
The change of the perspective on the collision model through the application of successive interference cancellation has led to \emph{coded random access}, an innovative approach superior to legacy SA. We have shown that the coded random access is tightly related to codes on graphs and we have presented several protocol variants. Considering that the ALOHA approach dominated during the last four decades, we believe that the coded random access opens new grounds for designing communication systems that should embrace a massive number of M2M devices. Finally, we note that principles of the coded random access can be combined with any MUD technique, i.e., they are not restricted to the simple chain of single-user detections' scenario assumed in the paper.

\bibliographystyle{IEEETran}

\begin{thebibliography}{10}
\providecommand{\url}[1]{#1}
\csname url@samestyle\endcsname
\providecommand{\newblock}{\relax}
\providecommand{\bibinfo}[2]{#2}
\providecommand{\BIBentrySTDinterwordspacing}{\spaceskip=0pt\relax}
\providecommand{\BIBentryALTinterwordstretchfactor}{4}
\providecommand{\BIBentryALTinterwordspacing}{\spaceskip=\fontdimen2\font plus
\BIBentryALTinterwordstretchfactor\fontdimen3\font minus
  \fontdimen4\font\relax}
\providecommand{\BIBforeignlanguage}[2]{{%
\expandafter\ifx\csname l@#1\endcsname\relax
\typeout{** WARNING: IEEEtran.bst: No hyphenation pattern has been}%
\typeout{** loaded for the language `#1'. Using the pattern for}%
\typeout{** the default language instead.}%
\else
\language=\csname l@#1\endcsname
\fi
#2}}
\providecommand{\BIBdecl}{\relax}
\BIBdecl

\bibitem{R1975}
L.~G. Roberts, ``ALOHA packet system with and without slots and capture,''
  \emph{SIGCOMM Comput. Commun. Rev.}, vol.~5, no.~2, pp. 28--42, Apr. 1975.

\bibitem{OIN1977}
H.~Okada, Y.~Igarashi, and Y.~Nakanishi, ``Analysis and application of framed
  {ALOHA} channel in satellite packet switching networks - {FADRA} method,''
  \emph{Electronics and Communications in Japan}, vol.~60, pp. 60--72, Aug.
  1977.
  
 \bibitem{ZZ2012}
A.~Zanella and M.~Zorzi, ``{T}heoretical analysis of the capture
  probability in wireless systems with multiple packet reception
  capabilities,'' \emph{IEEE Trans. Commun.}, vol.~60, no.~4, pp. 1058--1071,
  Apr. 2012.
  
\bibitem{TDN2011}
A.S.~Tehrani, A.G. Dimakis, and M.J. Neely, 
``SigSag: Iterative detection through soft message-passing,'' \emph{IEEE J. Sel. Topics Signal Process.}, 
vol.~5, no.~8, pp.~1512--1523, Dec. 2011.

\bibitem{GGW2013}
J.~Goseling, M.~Gastpar, and J.~Weber, ``Physical-layer network coding on the
  random access channel,'' in \emph{Proc. 2013 IEEE Int. Symp. Inf. Theory},
  Jul. 2013, pp. 2339--2343.

\bibitem{CGH2007}
E.~Casini, R.~{De Gaudenzi}, and O.~{del Rio Herrero}, ``{C}ontention
  resolution diversity slotted {ALOHA} {(CRDSA)}: {A}n enhanced
  random access scheme for satellite access packet networks,''
  \emph{IEEE Trans. Wireless Commun.}, vol.~6, no.~4, pp. 1408--1419, Apr.
  2007.

\bibitem{L2011}
G.~Liva, ``{G}raph-based analysis and optimization of contention
 resolution diversity slotted {ALOHA},'' \emph{IEEE Trans. Commun.},
  vol.~59, no.~2, pp. 477--487, Feb. 2011.

\bibitem{LMS1998}
M.~G. Luby, M.~Mitzenmacher, and A.~Shokrollahi, ``{A}nalysis of random
  processes via {A}nd-{O}r tree evaluation,'' in \emph{Proc. the 9th
  ACM-SIAM SODA}, San Francisco, CA, USA, Jan. 1998.

\bibitem{PLC2011b}
E.~Paolini, G.~Liva, and M.~Chiani, ``{G}raph-based random access for the
  collision channel without feed-back: {C}apacity bound,'' in
  \emph{Proc. IEEE GLOBECOM 2011}, Houston, TX, USA, Dec. 2011.

\bibitem{LPLC2012}
G.~Liva, E.~Paolini, M.~Lentmaier, and M.~Chiani, ``{S}patially-coupled
  random access on graphs,'' in \emph{Proc. IEEE ISIT 2012}, Boston,
  MA, USA, Jul. 2012.

\bibitem{SPV2012}
C.~Stefanovic, P.~Popovski, and D.~Vukobratovic, ``{F}rameless {ALOHA} protocol
  for wireless {networks},'' \emph{IEEE Comm. Letters}, vol.~16, no.~12, pp.
  2087--2090, Dec. 2012.

\bibitem{BLMR1998}
J.~Byers, M.~Luby, M.~Mitzenmacher, and A.~Rege, ``{A} digital fountain
  approach to reliable distribution of bulk data,'' in \emph{Proc.
  ACM SIGCOMM 1998}, Vancouver, BC, Canada, Sep. 1998.

\bibitem{SP2013}
C.~Stefanovic and P.~Popovski, ``{ALOHA} random access that operates as a
  rateless code,'' \emph{IEEE Trans. Commun.}, vol.~61, no.~11, pp.
  4653--4662, Nov. 2013.

\bibitem{STPP2013}
C.~Stefanovic, K.~F. Trilingsgaard, N.~K. Pratas, and P.~Popovski, ``{J}oint
  estimation and contention-resolution protocol for wireless random
  access,'' in \emph{Proc. IEEE ICC 2013}, Budapest, Hungary, Jun. 2013.

\bibitem{3GPPTS36.321}
\emph{{M}edium {A}ccess {C}ontrol ({MAC}) protocol specification}, 3GPP TS
  36.321.

\end{thebibliography}

\newpage

\newpage

\newpage

\newpage

\end{document}